\documentstyle[aps,prl,psfig,twocolumn]{revtex} 
\begin{document}
\title{
Jet quenching and event-wise mean-$p_{t}$ fluctuations
in Au-Au collisions
at $\sqrt{s_{NN}} = $ 200 GeV in Hijing-1.37 }

\author{
Qingjun Liu\thanks{On leave from Chinese Academy of Sciences, P.R. China}
and Thomas A. Trainor\\
CENPA, Box 354290, University of Washington, Seattle, WA 98195}
\date{\today}
\maketitle
\begin{abstract}
Based on Hijing-1.37 simulations, event-wise mean-$p_{t}$ fluctuations in Au-Au collisions at $\sqrt{s_{NN}}\:=\:200$ GeV are studied with graphical and numerical methods. This study shows that jets/minijets could represent a substantial source of mean-$p_t$ fluctuations, and jet quenching could reduce these fluctuations significantly. Mean-$p_t$ fluctuation measurements are a promising probe of correlation structure produced in the early stages of central Au-Au collisions at RHIC energies.\\ \\
\noindent PACS number: 25.75.Ld\\ \\
{\rm Keywords}: jet quenching, mean-$p_{t}$ fluctuations, Hijing, heavy ion collision
\end{abstract}
\section{Introduction}

QCD lattice gauge calculations \cite{QCD}  suggest that a phase transition from hadronic nuclear matter to a color-deconfined medium or quark-gluon plasma (QGP) may occur in relativistic heavy ion collisions at RHIC. Measurements of multiplicity and  transverse-momentum fluctuations have been proposed to establish the existence\cite{Rstock,Gbaym} and study the properties\cite{Stodolsky,Shuryak,SRS,Berdnikov,droplets}
of a QGP. It was suggested that in case of a second-order phase transition fluctuations may be reduced dramatically \cite{Stodolsky,Shuryak,SRS,Berdnikov}, whereas for a first-order phase transition large fluctuations may occur due to localized QGP droplet formation \cite{droplets}, or more gradual density or temperature fluctuations may occur. The past two years have seen significant progress in the number and quality of preliminary fluctuation and correlation measurements at both SPS\cite{Na45,Na49} and RHIC\cite{Phenix}. Further efforts are now required to draw quantitative conclusions about the phase transition and the physical characteristics of the QGP.

In this Letter we consider the possibility that a major source of event-wise mean-$p_t$ or $\langle p_t \rangle$ fluctuations in relativistic heavy ion collisions is jet/minijet production, and that these fluctuations can therefore be used to study the early collision process and the properties of the QGP, in a manner complementary to studies of {\em inclusive} event properties such as transverse momentum distributions \cite{Quench}.

According to \cite{pundits} minijets (semi-hard parton scattering with momentum transfers of a few GeV/c) should be copiously
produced in the initial stage of ultra-relativistic heavy ion collisions at RHIC energies. A QGP formed in the early stage of such collisions could act as a dissipative medium, causing jets/minijets to lose energy through induced gluon radiation \cite{Quench}, a process conventionally called jet quenching in the case of higher-$p_t$ partons. The properties of the dissipative medium should determine the specific energy loss of jets and minijets \cite{QuenDen}. We explore the role of fluctuation measurements in understanding the dissipation properties of a color-deconfined medium, using Hijing-1.37 \cite{Hijing} to study jet-quenching and jet/minijet contributions to $\langle p_t \rangle$ fluctuations.

\section{Hijing Event Types} 


Three sets of Hijing-1.37 events were generated for this
study: Hijing default with jet quenching on, with quenching switched off and with jet/minijet production switched off. Because jet quenching effects should be most significant in central collisions this study was applied to Hijing-1.37 events with impact parameter less than 3 fm, equivalent to the top 5\% most central collisions. 

Fluctuation analysis was applied to all generated charged particles  subject to the conditions that the pseudorapidity and transverse momentum satisfied $|\eta| < 1$ and 0.15 GeV/c $< p_{t} <$2 GeV/c. The total number of events, the mean $\hat p_t$  and variance $\sigma_{\hat p_t}$  of the charged-particle inclusive $p_t$ distributions, and the mean $\bar{n}$ and variance $\sigma_{n}$  of the charged-particle multiplicity distributions for the three types of central Hijing events are listed in Table~\ref{tbl}. We analyze $\langle p_t \rangle$ fluctuations for these three event classes using graphical and numerical methods.

\section{Graphical Analysis of Fluctuations}

Graphical analysis of $\langle p_{t} \rangle$ fluctuations directly compares the data distribution with a reference. The gamma distribution models statistically-independent particle emission \cite{Mjt}. Any difference between data and reference would imply that emitted particles are correlated and/or event-wise fluctuations in collision dynamics (such as minijets) are present. Effects of fluctuating event multiplicity in this method require folding the gamma distribution with a negative binomial distribution \cite{Mjt}. The reference distribution is formulated as:

\begin{equation} \label{sie}
f(\langle p_{t} \rangle)\:=\: \sum_{n_{min}}^{n_{max}} f_{NBD}(n,1/k,\bar n)f_{\Gamma}(\langle p_{t} \rangle,np,nb)
\end{equation}
where 
$f_{NBD}(n,1/k,\bar n)$ and $f_{\Gamma}(\langle p_{t} \rangle,np,nb)$ are the negative binomial distribution and n-fold gamma distribution respectively
\begin{eqnarray} \label{NBD}
f_{NBD}(n,1/k,\bar n) &=& \frac{(n+k-1)!}{n!(k-1)!}\frac{(\frac{\bar n}{k})^n}{(1+\frac{\bar n}{k})^{n+k}} \\ \nonumber
f_{\Gamma}(\langle p_{t} \rangle,np,nb) &=& \frac{nb}{\Gamma(np)}(nb\langle p_{t} \rangle)^{np-1} \exp^{-nb\langle p_{t} \rangle}
\end{eqnarray}
\noindent
The parameters in Eq. (\ref{NBD}) are defined as
$ p = {\hat p_{t}^2} / {\sigma_{\hat p_{t} }^2}$, $b = {\hat p_t} / {\sigma_{\hat p_t}^{2}}$ and 
 $ {\bar n} / {k} =  {\sigma_n^{2}} / {\bar n^{}}-{1}{} $, where $\hat p_t$ and $\sigma^2_{\hat p_t}$ are the mean and variance respectively of the inclusive $p_t$ distribution. This model is related to the central limit theorem result that the cumulants of a distribution of independent $n$-sample means from a fixed parent distribution are related to the cumulants of the parent by inverse powers of the sample number. The gamma distribution cumulants satisfy this relation, and the inclusive $p_t$ distribution is approximately described by a gamma distribution with $n=2$. The relations above then follow. The graphical comparison is important as a qualitative indicator of excess variance and of significant deviations from central limit behavior in cumulants higher than the second.

\subsection{Basic histograms}

\begin{figure}[h]
\centerline{\psfig{figure=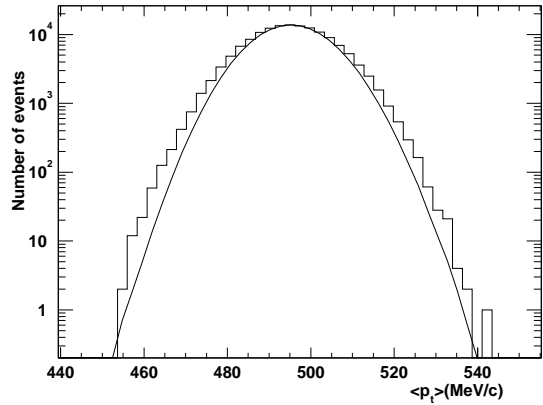,height=6.cm}}
\centerline{\psfig{figure=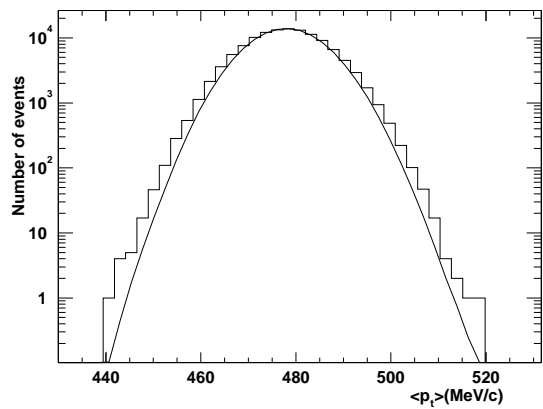,height=6.cm}}
\centerline{\psfig{figure=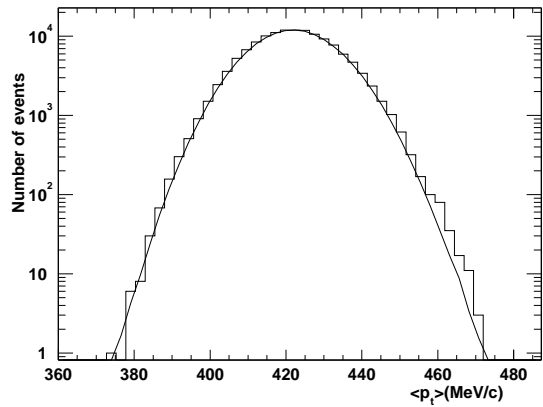,height=6.cm}}
\caption[]{Distributions of $\langle p_{t} \rangle$
for Au-Au collisions at $\sqrt{s_{NN}}$ = 200 GeV from Hijing-1.37 (histograms) compared with reference distributions assuming statistically-independent particle emission and calculated using Eq.~(\ref{sie}) (curves). The top panel shows results for jets-on quench-off events, the middle panel for jets-on quench-on events and the bottom panel is for jets-off events.}
\label{fig:mj}
\end{figure}

Distributions of $\langle p_t \rangle$ for each of three Hijing event types are shown in Fig.~\ref{fig:mj} as histograms together with reference distributions based on Eq.~(\ref{sie}) plotted as curves.
The reference distributions are normalized to have the same peak values as the corresponding data distributions to emphasize the width comparison. It can be seen from Fig.~\ref{fig:mj} that the
Hijing data histograms deviate substantially from the reference distributions, indicating that nonstatistical fluctuations are significant. Fig.~\ref{fig:mj} indicates that excess fluctuations in events with jets switched off are substantially less than those for the other two event classes.

Comparison of the top and bottom panels of Fig.~\ref{fig:mj} indicates qualitatively that jet production plays a significant role in $\langle p_t \rangle$ fluctuations in the Hijing model. To investigate the effect of jet quenching we can compare nonstatistical fluctuations in top and middle panels of Fig.~\ref{fig:mj}. However, with these histograms it is not possible to detect visually a significant difference in the magnitude of nonstatistical $\langle p_t \rangle$ fluctuations. We require a more differential approach.

\subsection{Difference histograms}

The graphical analysis can be made more sensitive by plotting the difference between data histogram and reference distribution, facilitating quantitative comparisons between Monte Carlo configurations and references. The normalized event-number difference is defined as
\begin{equation}
	\delta N / \sqrt{N} \:\equiv\: \frac{N_{data}\:-\:N_{reference}}{\sqrt{N_{data}}}
\end{equation}
The independent random variable $\sqrt{\bar {n}}\,(\langle p_{t} \rangle\:-\:\hat p_t) / \sigma_{\hat p_t}$ 
normalizes the distribution width relative to the inclusive width according to a central limit expectation. The result is a universal plotting format; the horizontal and vertical axis units are standard deviations.

\begin{figure}[h]
\centerline{\psfig{figure=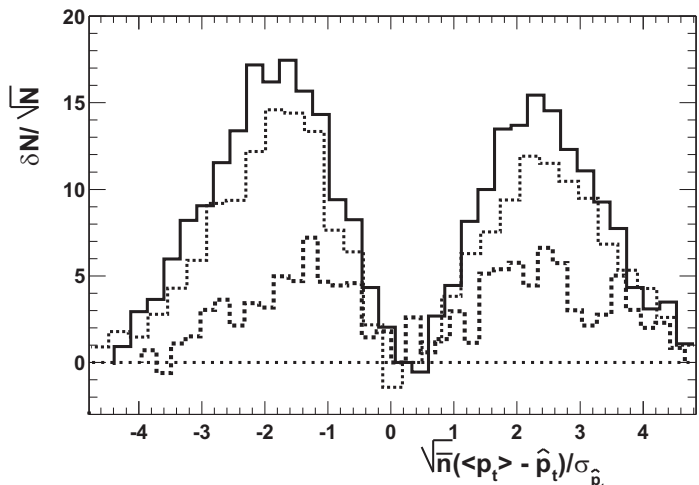,height=6.5cm}}
\caption{Normalized event-number difference $\delta N / \sqrt{N}$ {\rm vs} $\sqrt{\bar {n}}\,(\langle p_{t} \rangle\:-\:\hat p_t) / \sigma_{\hat p_t}$ for three types of Hijing-1.37 events, for Au-Au collisions at $\sqrt{s_{NN}}$ = 200 GeV. Solid line shows results for jets-on quench-off events, dashed line for jets-on quench-on events and dotted line for jets-off events. 
}
\label{fig:tr}
\end{figure}

Figure~\ref{fig:tr} shows results for the three Hijing-1.37 configurations. Significant deviations from zero reveal nonstatistical fluctuations in all three event types. The dramatic increase from jet-off events (lowest curve) to events with jets on is consistent with the qualitative difference in Fig. \ref{fig:mj}. A significant difference with and without jet quenching (upper two curves) is also now apparent. This difference indicates that according to the Hijing model transverse momentum fluctuations are sensitive to jet quenching, which may in turn reflect the properties of a dense colored medium or QGP formed in relativistic heavy ion collisions. We conclude that while the original graphical method is an indicator for fluctuations associated with jet production, the difference histograms are more sensitive to the effects of jet quenching.

\section{Numerical Analysis of Fluctuations}

Numerical analysis of nonstatistical fluctuations requires comparison of a variance from data with the variance of a reference representing only statistical fluctuations. The linear variance comparison measure $\Delta \sigma_{p_{t}}$ \cite{Trainor} was developed to isolate pure $\langle p_t \rangle$ fluctuations from effects of multiplicity fluctuations which may appear as a systematic bias in some statistical measures. 

Several quantities have been proposed to measure event-wise $\langle p_t \rangle$ fluctuations in relativistic heavy ion collisions:
$f_{p_t}$\cite{SRS}, $F_{p_{t}}$\cite{Phenix}, $\Phi_{p_{t}}$\cite{PhiPt}, and $\sigma^2_{p_{t},dynamical}$\cite{SigD}.
A detailed study of $\Phi_{p_{t}}$ is given in \cite{Trainor}. In the present analysis the closely-related $\Delta \sigma_{p_{t}}$ is applied to event-wise $\langle p_{t} \rangle$ fluctuations in Hijing-1.37 events. This {\em difference factor} for $\langle p_t \rangle$ fluctuations is defined as
\begin{equation}
	\Delta \sigma_{p_{t}}\:\equiv\: \left\{ \overline{n( \langle p_{t} \rangle \:-\: \hat p_t)^{2}}\:-\: \sigma_{\hat p_t}^{2} \right\} / {2\sigma_{\hat p_t}}
\end{equation}
where the overline means average over all events in a data set. Values of $\Delta \sigma_{p_{t}}$ for the three Hijing event types are included in Table~\ref{tbl}. Consistent with Fig. \ref{fig:mj} the results show very significant contributions from jets, and they also show that fluctuations are significantly reduced when jet quenching is applied, consistent with Fig. \ref{fig:tr}. 

As expected, numerical analysis of $\langle p_t \rangle$ fluctuations reveals more precisely than graphical analysis that jet/minijet production in the Hijing model contributes significantly to nonstatistical event-wise $\langle p_{t} \rangle$ fluctuations. The increase in fluctuations for jets on compared to jets off is a factor 3-5. The effect of jet quenching is to reduced $\Delta \sigma_{p_{t}}$ by about 30\% relative to no quenching, 20 times the statistical error. These results suggest that properties of the QGP or dissipative medium formed in the initial stage of heavy ion collisions may be studied precisely with numerical analysis of event-wise $\langle p_{t} \rangle$ fluctuations. 

\section{Centrality Dependence}

However, we observe an excess of binary-collision scaling in Hijing compared to RHIC data which affects the interpretation of this fluctuation analysis. Hijing  particle
production increases with number of binary collisions 
\begin{table}[tb]
\caption{The total number of events, mean $\hat p_t$ and
variance $\sigma_{\hat p_t}$ of inclusive charged-particle
transverse-momentum distributions, the mean  $\bar{n}$ and variance $\sigma_{n}$ of charged-particle multiplicity distributions, and the event-wise mean-$p_{t}$ excess fluctuations $\Delta \sigma_{p_{t}}$
for Hijing-1.37 events --- 5\% central (b $ < 3$ fm) Au-Au collisions at $\sqrt{s_{NN}}~=~200~$ GeV, $|\eta|<1$. Errors are statistical only. \label{tbl}}
\begin{tabular}{|c||c|c|c|}
event class & no jets & jets - quench on & jets - quench off \\
\hline
event total & 134868 & 121671 & 146241 \\
\hline \hline
$\bar{n}$& 431 & 1387 & 1159 \\
\hline
$\sigma_n$ & 30 & 130 & 100 \\
\hline
$\hat p_t$(MeV/c) & 424 & 478 & 496 \\
\hline 
$\sigma_{\hat p_t}$(MeV/c) & 227 & 283 & 307 \\
\hline \hline
$\Delta \sigma_{p_{t}}$(MeV/c) & $7.2 \pm 0.3$ & $24.9 \pm 0.5$ & $34.5 \pm 0.5$ \\
\end{tabular}
\end{table}
(centrality) at twice the rate observed in RHIC Au-Au collisions, as indicated by the multiplicity entries in Table \ref{tbl}. The jets-off multiplicity reflects soft processes, whereas 
for the same number of participant pairs the jets-on multiplicities, which include hard-component scaling with binary collisions, represent  a 2.5-3-fold increase over participant scaling for central events. For RHIC data we observe a corresponding 1.5-fold increase \cite{phobos}. The fraction of particle production from hard processes in Hijing events is thus about twice that observed for RHIC events. We find that $\langle p_t \rangle$ fluctuations in Hijing (jets on) are dominated by the hard component, by a factor 5. It may be advisable therefore to reduce by a factor of two the $\langle p_t \rangle$ fluctuations attributed by this study to jets/minijets in Hijing in comparisons with RHIC collisions.

\section{Hadronic Rescattering}

Hadron rescattering between chemical and kinetic de-coupling might attenuate nonstatistical fluctuations generated in the early stage of the collision, which could then be confused with effects of dissipation in the prehadronic medium. Recent Monte Carlo calculations suggest that hadronic rescattering may be negligible \cite{Gbaym,Kwerner}: the time for rescattering may be so brief \cite{Jrafelski,Thermal} that fluctuation attenuation is not significant. Nonstatistical $\langle p_{t} \rangle$ fluctuations due to jets/minijets would thus survive hadronic rescattering unscathed. This is an important point for interpreting attenuation of  fluctuations observed in the final state. Reduced $\langle p_{t} \rangle$ fluctuations as reported in this Letter would then be an unambiguous indication of prehadronic dissipation (minijet/jet quenching).

\section{Conclusions}

In summary, event-wise $\langle p_{t} \rangle$ fluctuations in
central Au-Au collisions at $\sqrt{s_{NN}}\:=\:200$ GeV have been studied with Hijing-1.37 using graphical and numerical methods. The study shows that jet/minijet production in the initial stage
of heavy ion collisions may contribute substantially to nonstatistical $\langle p_{t} \rangle$ fluctuations. We have also shown that in the Hijing model $\langle p_{t} \rangle$ fluctuations are sensitive to jet quenching, thus suggesting that fluctuation analysis could serve as a probe of dissipation properties of a dense colored medium formed in the initial stages of ultrarelativistic heavy ion collisions. Studies of fluctuation centrality dependence in models and data may help clarify questions on hadronic dissipation and Hijing binary-collisions scaling raised by this study.

We appreciate helpful discussions with J.G. Reid (University of Washington). This work was supported in part by USDOE contract DE-FG03-97ER41020.
 


\begin{thebibliography}{9}
\bibitem{QCD} J.~C.~Collins and M.~Perry, Phys. Lett. {\bf 34} (1975) 1353;
	B.~Friedman and L.~McLerran, Phys. Rev. {\bf D 17} (1978);
	G.~Baym and S.~A.~Chin, Phys. Lett. {\bf B 62} (1986) 241;
	E. Laemann, Nucl. Phys. {\bf A 610} (1996) 1c.
\bibitem{Rstock} R. Stock, Nucl. Phys. {\bf A 661} (1999) 282.
\bibitem{Gbaym} G. Baym and H. Heiselberg, Phys. Lett. {\bf B 469} (1999) 7;
\bibitem{Stodolsky} L. Stodolsky, Phys. Rev. Lett. {\bf 75} (1995) 1044.
\bibitem{Shuryak} E.~V.~Shuryak, Phys. Lett. {\bf B 430} (1998) 9;
	M. Stephanov, K. Rajagopal, and E. Shuryak,
	Phys. Rev. Lett. {\bf 81} (1998) 4816.
\bibitem{SRS}M. Stephanov, K. Rajagopal, and E. Shuryak,
	Phys. Rev. {\bf D 60} (1999) 114028.
\bibitem{Berdnikov}
	B. Berdnikov and K. Rajagopal, Phys. Rev. {\bf D 61} (2000) 105017;
	K. Rajagopal, Nucl. Phys. {\bf A 680} (2000) 211.
\bibitem{droplets} L. Van Hove, Z. Phys. {\bf C 21} (1984) 93;
	E.E. Zabrodin, L.P. Csernai, J.I. Kapusta, G. Kluge,
	Nucl. Phys. {\bf A 566} (1994) 407c;
	J.I. Kapusta, A.P. Vischer, Phys. Rev. {\bf C 52} (1995) 2725.
\bibitem{Na45} CERES/NA45 Collab., H. Appelshauser et al.,
	Nucl. Phys. {\bf A 698} (2002) 253c.
\bibitem{Na49} NA49 Collab., H. Appelshauser et al.,
	Phys. Lett. {\bf B 459} (1999) 679686.
\bibitem{Phenix} Phenix Collab., K. Adcox et al., Phys. Rev. {\bf C 66} (2002) 024901.
\bibitem{Quench} M. Gyulassy and M. Plumer, Phys. Lett. {\bf B 243} (1990) 432;
	X.N. Wang and M. Gyulassy, Phys. Rev. Lett. {\bf 68} (1992) 1480.
\bibitem{pundits} K. Kajantie, P.V. Landshoff, J. Lindfors, Phys. Rev. Lett. {\bf 59} (1987) 2527.
\bibitem{QuenDen}
	M. Gyulassy and X.N. Wang, Nucl. Phys. {\bf B 420} (1994) 583; R. Baier,
	Y.L. Dokshitzer, S. Peigne and D. Schiff, Phys. Lett. {\bf B 345} (1995) 277.
\bibitem{Hijing}
	M. Gyulassy and X.N. Wang, Comp. Phys. Comm. {\bf 83} (1994) 307;
	Phys. Rev. D 44 (1991) 3501.
\bibitem{Mjt} M.J. Tannenbaum, Phys. Lett. {\bf B 498} (2000) 29;
\bibitem{Trainor} T.A. Trainor, {\it hep-ph/0001148}.
\bibitem{PhiPt} M. Gazdzicki, St. Mrowczynski, Z. Phys. {\bf C 54} (1992) 127.
\bibitem{SigD} S.A. Voloshin, V. Koch, H.G. Ritter,
	Phys. Rev. {\bf C 60} (1999) 024901.
\bibitem{phobos} B.B. Back {\em et al.}, Phys Rev. {\bf C 65} (2002) 061901.
\bibitem{Kwerner} H. Heiselberg, Phys. Rept. {\bf 351} (2001) 161;
Phys. Rev. {\bf C 66} (2002) 044902.
\bibitem{Jrafelski}
	J. Rafelski and J. Letessier, Phys. Rev. Let. {\bf 85} (2000) 4695;
	G. Torreri and J. Rafelski, New J. Phys. {\bf 3} (2001)~~12.
\bibitem{Thermal}
	W. Florkowski and W. Broniowski, {\it nucl-th/0208061};
   	W. Broniowski and W. Florkowski, Phys. Rev. Lett. {\bf 87} (2001) 272302.
\end{thebibliography}
\end{document}